# Photo-switchable nanoripples in Ti$_3$C$_2$$T_x$ MXene

*Mikhail Volkov[1,2,*], Elena Willinger[3,*], Denis A. Kuznetsov[3], Christoph R. Müller[3], Alexey Fedorov[3] and Peter Baum[1,2]*

[1] University of Konstanz, Universitätsstraße 10, 78457 Konstanz, Germany

[2] Ludwig-Maximilians-Universität München, Am Coulombwall 1, 85748 Garching, Germany

[3] Department of Mechanical and Process Engineering, ETH Zürich, Leonhardstrasse 21, 8092 Zürich, Switzerland

**\*Authors to whom correspondence should be addressed:**

mikhail.volkov@uni.kn

elenawi@ethz.ch

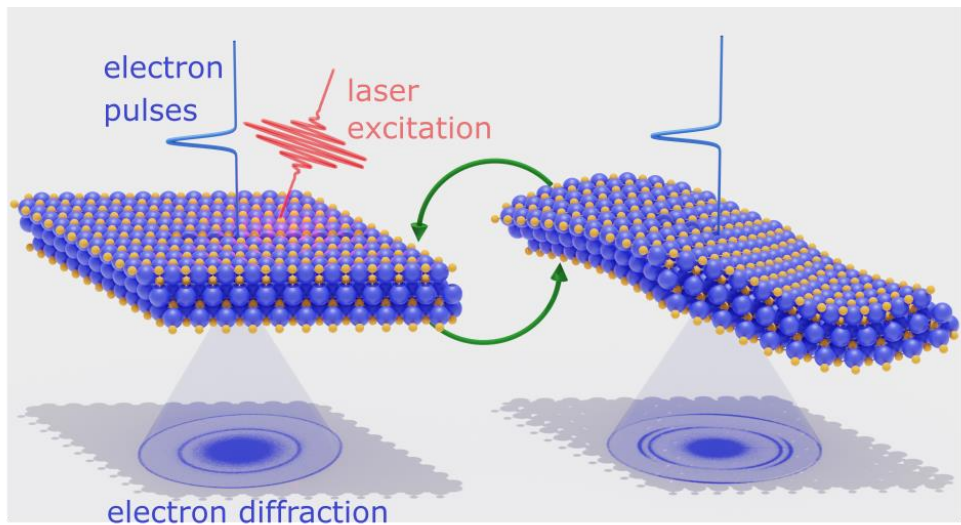

**ABSTRACT:** MXenes are two-dimensional materials with a rich set of remarkable chemical and electromagnetic properties, the latter including saturable absorption and intense surface plasmon resonances. To fully harness the functionality of MXenes for applications in optics, electronics and sensing, it is important to understand the interaction of light with MXenes on atomic and femtosecond dimensions. Here, we use ultrafast electron diffraction and high-resolution electron microscopy to investigate the laser-induced structural dynamics of $Ti_3C_2T_x$ nanosheets. We find an exceptionally fast lattice response with an electron-phonon coupling time of 230 femtoseconds. Repetitive femtosecond laser excitation transforms $Ti_3C_2T_x$ through a structural transition into a metamaterial with deeply sub-wavelength nanoripples that are aligned with the laser polarization. By a further laser illumination, the material is reversibly photo-switchable between a flat and rippled morphology. The resulting nanostructured MXene metamaterial with directional nanoripples is expected to exhibit an anisotropic optical and electronic response as well as an enhanced chemical activity that can be switched on and off by light.

**MAIN TEXT**

MXenes are a large and rapidly-expanding family of two-dimensional transition metal carbides, nitrides and carbonitrides in form of few-atom-thick nanosheets. The general chemical formula is $M_{n+1}X_nT_x$, where n = 1-3, M is an early transition metal, X is carbon or nitrogen and $T$ is a surface termination group, such as O, OH or F.[1,2] MXenes exhibit a plethora of unique chemical, mechanical, optical and electronic properties that find applications in chemical and mechanical sensing, energy storage, electronics and nonlinear optics, to name a few.[3] Insight from numerous studies has paved the way to rationalize, predict, engineer and exploit those properties of MXenes for various applications[3–7]. However, MXenes with a controlled nanostructure, for example with laser-induced periodic surface structures (LIPSS)[8], are currently not available, despite potential advantages of such materials for photocatalysis, antibiotic activity, charge storage or surface-enhanced Raman scattering. [9–14]

Among the broad class of available MXenes, $Ti_3C_2T_x$ has been synthesized first[15] and is to date the most widely investigated material, both theoretically and experimentally. The 2D nature and metallic conductivity of $Ti_3C_2T_x$ give rise to peculiar optical properties and light-driven phenomena. For example, $Ti_3C_2T_x$ is highly effective in shielding electromagnetic interference[16] and its wide optical absorption spectrum allows for efficient light-to-heat conversion.[17] Irradiation in ambient air or water solution has been shown to accelerate degradation of $Ti_3C_2T_x$ into $TiO_2$ and amorphous titanium carbide in the laser-treated regions,[18,19] and focused femtosecond laser beams can thereby create flexible electronic circuits.[20,21] At high light intensity, $Ti_3C_2T_x$ exhibits broadband saturable absorption that enables effective laser mode-locking and other nonlinear photonics applications.[22–24] The majority of such effects, however, are currently discussed from a rather phenomenological and applied perspective, and a time-resolved, atomic-scale description of light-matter interaction in MXenes is not yet fully developed.

Here we use ultrafast electron diffraction as a direct probe of structural dynamics in laser-excited $Ti_3C_2T_x$ with femtosecond and atomic resolutions in time and space,[25,26] complemented by an atomic-resolution transmission electron microscopy study of the laser-processed materials. We find that $Ti_3C_2T_x$ exhibits a very fast, 230-fs lattice response to laser excitation. Repetitive laser irradiation induces 20-nm periodic ripples that are aligned with the electric field of the laser light. The nanoripples are reversibly flattened out during further laser irradiation, rendering this morphological transformation photo-switchable.

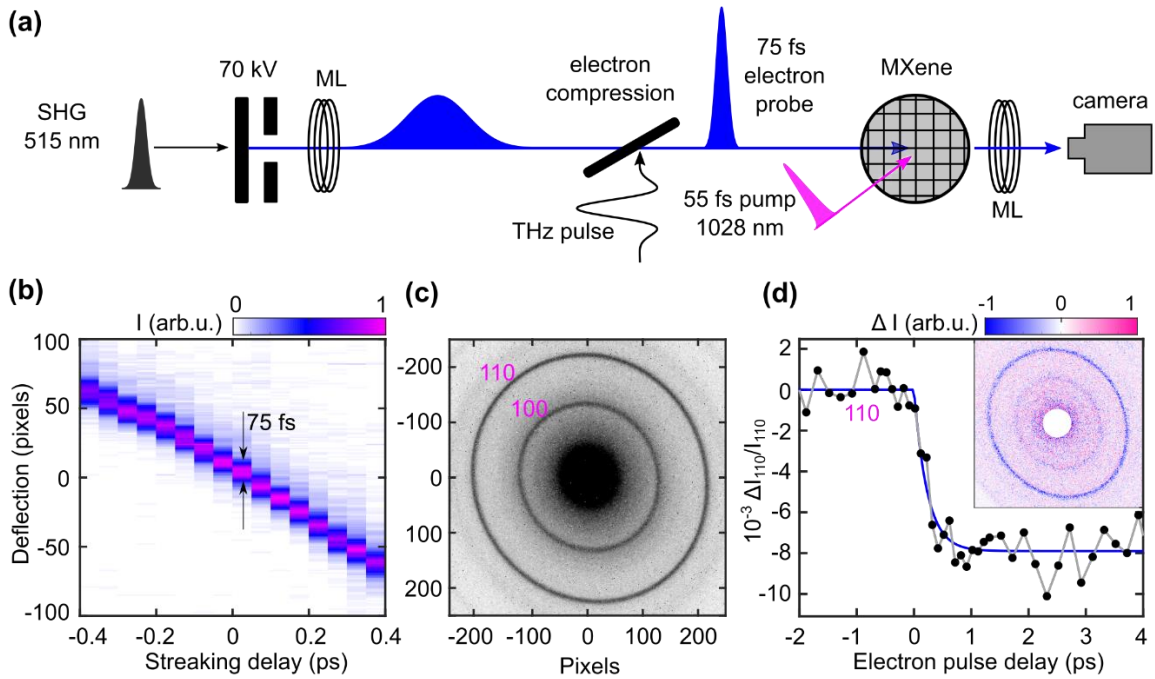

**Figure 1**. **(a)** Experimental layout of the ultrafast electron diffraction experiments with terahertz electron compression. A laser source synchronously drives electron pulse generation via its second harmonic (SHG, 515 nm wavelength), produces THz pulses for compression and streaking, and provides 55-fs pump pulses for sample excitation. Electron pulses at an energy of 70 keV and with about 8 electrons/pulse are focused with a magnetic lens (ML) and compressed down to a pulse duration of 75 fs with a THz pulse on an aluminum foil (black). A bow-tie resonator (not depicted) provides temporal characterization of the pulses via electron streaking. A free-standing film of $Ti_3C_2T_x$ on a copper mesh is pumped by a near-infrared laser pulse at 1028 nm wavelength. **(b)** THz streaking of the compressed electron pulse reveals a slope of 0.17 pixels/fs (black line) and an electron pulse duration of 75 fs. **(c)** MXene diffraction pattern. Two intense diffraction rings are observed at 0.39 Å$^{-1}$ and 0.67 Å$^{-1}$, corresponding to (100) and (110) indexes. **(d)** Ultrafast Debye-Waller effect in MXene. The intensity drop of the (110) diffraction ring occurs within 230 fs. Black dots show experimental data; blue curve shows a fit with a response function $e^{-t/\tau}$, $\tau$ = 230 fs.

## RESULTS

**Femtosecond electron diffraction**

Figure 1a depicts our femtosecond electron diffraction experiment.[27,28] Ultrashort electron pulses without space charge effects[29] are generated at a central energy of 70 keV by two-photon laser photoemission[30] and subsequent electrostatic acceleration. Single-cycle terahertz radiation is used to compress the electron pulses[27] down to a duration of 75 fs. Figure 1b shows terahertz streaking data from which we directly measure the electron pulse duration in the experiment. The specimen is a freestanding film of $Ti_3C_2T_x$ flakes that are prepared by chemical exfoliation (Methods) and then deposited onto a support grid with a mesh spacing of 300 μm. The sample is excited by femtosecond laser pulses with a center wavelength of 1028 nm and a pulse duration of 55 fs, yielding in total a sub-100-fs time resolution of our ultrafast electron diffraction experiment.

Figure 1c shows the static electron diffraction pattern of the $Ti_3C_2T_x$ flakes as observed with our femtosecond pulses. The two most prominent rings correspond to Bragg indices of (100) and (110) with reciprocal lattice vectors of 0.39 Å$^{-1}$ and 0.67 Å$^{-1}$, respectively. Slight deviations from a circular shape are due to distortions in the large magnetic lenses after the electron-sample interaction. The presence of mainly (100) and (110) diffraction rings, their sharpness and high intensity indicate that all flakes are lying flat on the support mesh with the preferred orientation along the c-axis. The intensity of the both diffraction rings shows a homogeneous azimuthal distribution, indicating random rotational orientation of the stacked 2D flakes.

Femtosecond laser excitation with a fluence of 0.9 mJ/cm$^2$ induces a decrease of the measured diffraction intensities in the rings as a consequence of the ultrafast Debye-Waller effect. The inset of Figure 1d shows the difference of the diffraction patterns before and after time zero, exhibiting an isotropic intensity drop of about 0.8%. Ring diameters and widths remain constant within the sensitivity of our experiment. Figure 1d shows the integrated electron counts in the (110) diffraction ring as a function of the pump-probe delay. We determine the response time from a fit with an exponential response function (blue solid line) and obtain an electron-phonon coupling rate of 230±105 fs at 95% confidence. We see no further dynamics of the diffraction pattern in the following 30 picoseconds.

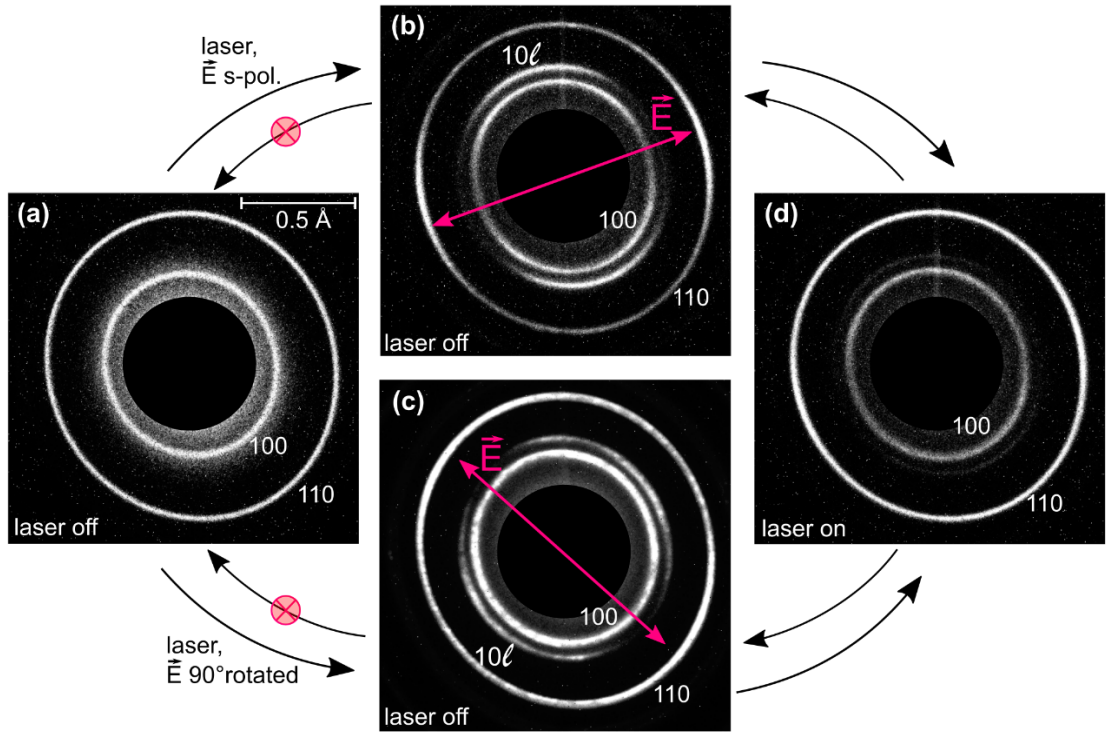

**Figure 2.** Laser-induced symmetry breaking. **(a)** Electron diffraction pattern from pristine $Ti_3C_2T_x$. **(b)**, **(c)** Electron diffraction after laser processing with orthogonal polarizations (magenta arrows). Data is taken at room temperature with the laser off. **(d)** Electron diffraction from a laser-processed sample while the laser in on. The processing is irreversible (black arrows with red crosses) but the switching is reversible (black arrows).

**Field-induced nanoripple formation**

At a higher laser fluence, above 1 mJ/cm², we observe a permanent structural transition that manifests as new diffraction arcs that appear at an inverse atomic distance of 0.47 Å$^{-1}$ and break the azimuthal symmetry of the ring pattern. Figure 2b shows such a diffraction pattern, obtained by illuminating the specimen with s-polarized femtosecond laser pulses at an incidence angle of 25° with a peak fluence of 1.6 mJ/cm² for 10 seconds and then letting the sample relax back to room temperature. Notably, the broken symmetry aligns with the electric field direction of the femtosecond laser (Figure 2b, magenta arrows). Figure 2c shows the diffraction pattern of a fresh sample that is illuminated with femtosecond laser pulses of perpendicular polarization. We see a 90° rotation of the symmetry-breaking additional diffraction arcs. Rotation of the polarization by 45° produces a 45° arc rotation. Once processed, the samples retain their arc orientation and cannot be switched into other arc orientations by further illumination attempts.

To assess quantitatively the transition threshold, that is, the minimum optical power that is required to produce the arcs, we gradually increase the laser fluence that impinges on a fresh

Ti$_3$C$_2$T$_x$ specimen. After application of each laser fluence, we let the specimen relax to room temperature and then measure the intensity of the diffraction rings. We find that the symmetry-breaking arcs appear at a laser fluence of 1 mJ/cm$^2$, corresponding to a laser power of 9 mW, but not at lower laser powers even for longer illumination times (Figure S3a). The same experiments at twice the laser repetition rate and twice the average power produce the same per-pulse fluence threshold of ca. 1 mJ/cm$^2$. This result indicates that the transformation is non-thermal but must be directly related to the high peak intensity of the femtosecond pulses. From sample to sample, the fluence threshold varies by about ±0.5 mJ/cm$^2$. An increase of the laser fluence to ca. 130% of the transition threshold initiates a gradual evaporation of the flakes, observed as an irreversible decrease of electron intensity in all the diffraction rings (Figure S3a).

While a laser-processed specimen is under laser illumination above the formation threshold but below the damage threshold, the symmetry-breaking arcs diminish and reduce their intensity almost to zero (Figure 2d). At the same time, the intensity of the (110) ring increases to a higher intensity than before (Figure S3b), indicating a higher degree of orientation with respect to the $c$ axis. When switching the light off, the electron diffraction returns to the original symmetry-broken pattern that was created by the initial processing. More than 50 reversible phase changes are observed without a measurable degradation of the specimen (Figure S3b).

**Selected-area electron diffraction (SAED) and high-resolution transmission electron microscopy (HRTEM)**

In order to understand the laser-induced structural transformation, we report a transmission electron microscopy (TEM) analysis of the pristine and the laser-beam processed Ti$_3$C$_2$T$_x$ films. Figures 3a and 3b show selected-area electron diffraction (SAED) patterns of the pristine and the laser-processed Ti$_3$C$_2$T$_x$ films, respectively. The probed area has a dimeter of ca. 1.3 μm. Both diffraction patterns reproduce the femtosecond results of Figure 2. The pattern of the fresh sample is dominated by (100) and (110) rings, reflecting the preferential orientation of the two-dimensional Ti$_3$C$_2$T$_x$ films with alignment to the $c$ axis (see above). In contrast to the fresh material, the SAED pattern of the laser-processed Ti$_3$C$_2$T$_x$ films contains additional, relatively intense diffraction arcs near the (100) ring, as already seen in the femtosecond diffraction data (Figure 2). Notably, thermal (non-optical) heating up to 500° C with subsequent cooling does not produce any arcs (Figure S4). Rotational averaging of the two-dimensional (2D) diffraction pattern is used to generate one-dimensional (1D) diffraction data for further analysis (Figure 3d). A comparison of the two 1D diffraction data shows the re-distribution of intensity and growth of the arc feature in the laser-processed sample. The decrease of the (100) and the (110) peak intensities indicates a reduced effect of the preferential orientation as compared to the pristine Ti$_3$C$_2$T$_x$ film.

Additional SAED patterns are recorded as a function of tilt angle of the $Ti_3C_2T_x$ film with respect to the incident electron beam (Figure S2). We observe that the intensity of the (110) peak decreases while the intensity of diffraction arcs grows with increasing the tilt angle. A comparison of the SAED patterns of the laser-processed and pristine $Ti_3C_2T_x$ films tilted by −15° shows similarities in the diffraction intensities (Figures 3b, 3c, 3e). Therefore, both materials possess a similar crystallographic long-range order. The new diffraction arcs can be assigned to the (10$\ell$) diffraction planes (Figure 3f, magenta lines). The orientation of the (10$\ell$) planes with respect to the crystallographic $c$-axis inside the MXene structure is shown in Figure 3f. A similar arc formation effect is observed for example in highly-ordered pyroelectric graphite under a 20-degree tilt[31] or in thin-films with a lamellar or fibrous texture.[32]

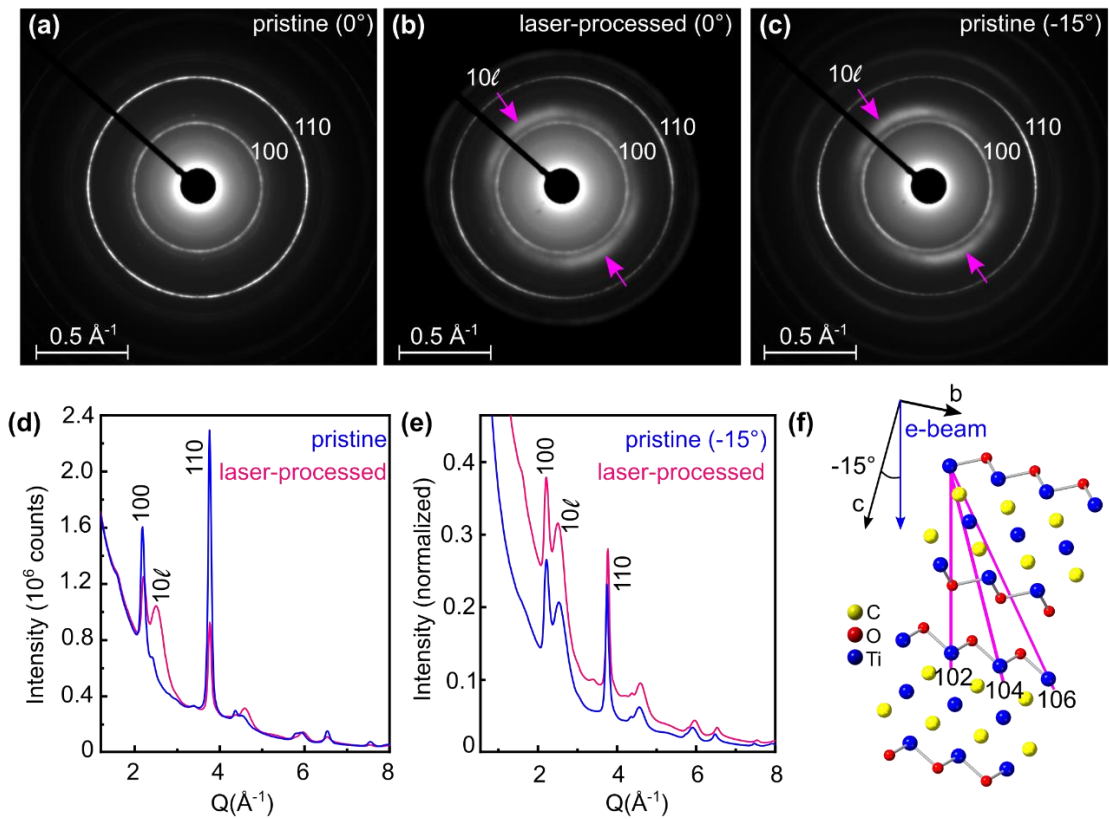

**Figure 3.** TEM analysis of pristine and laser-processed $Ti_3C_2T_x$ films. **(a)** SAED pattern of the pristine film with a 0° tilt angle relative to the e-beam. **(b)** SAED pattern of the laser-processed material. **(c)** SAED pattern of the pristine sample recorded at −15° tilt. **(d)** Comparison of the 1D SAED data of the pristine (blue) and laser-processed (magenta) materials. **(e)** Comparison of the 1D SAED data of the laser-processed (magenta) and the pristine (blue) $Ti_3C_2T_x$ films at −15° tilt. **(f)** Illustration of the (10$\ell$) planes, $\ell = \{2,4,6\}$, in a tilted MXene stack.

Since the MXene films consist of stacks of 2D nanosheets that do not feature a high degree of crystallographic order in the stacking direction, we use the atomic electron pair distribution function (ePDF) for a quantitative analysis of the local atomic structures of pristine and laser-processed $Ti_3C_2T_x$ films. Figure 4a shows the results in form of reduced structure factors, F(Q). We see that the data from the pristine (blue) and laser-processed specimen (magenta) are very similar (Figure 4a). In Figure 4b, the corresponding ePDF data (G(*r*) functions) are presented as light grey circles while the blue and magenta curves correspond to the refinements of the pristine film and a laser-processed specimen, respectively, using a model based on the reported $Ti_3C_2T_x$ structure.[33] The black lines denote the residuals of the fits. While the structure refinement of the pristine $Ti_3C_2T_x$ film (blue) provides an acceptable fit to the data with an agreement factor $R_w$ = 28%, the ePDF fit of the laser-processed MXene (magenta) shows a poorer agreement ($R_w$ = 39%, Figure 4b). However, the fit of a laser-processed specimen is reasonable at larger atomic distances above ca. 3Å, indicating a good match to the $Ti_3C_2T_x$ structure in the medium-range order. A clear misfit is only observed in the region of low interatomic distances (compare the top and bottom black residual data in Figure 4b). There, the laser-processed $Ti_3C_2T_x$ specimen features an additional peak at an interatomic distance of ca. 1.7 Å that cannot be accounted for by the reported $Ti_3C_2T_x$ structural model.[34] The peak is most likely due to shortened Ti-O/F/C bond(s) after the laser exposure (the enlarged region in Figure 4c). This relatively short bond length is characteristic for [$TiO_4$] tetrahedra that are found in some titanates (for example in $CsAlTiO_4$, ICSD: 65707, Ti–O bond length is 1.75 Å). Therefore, it appears likely that a loss or reconfiguration of terminal $T_x$ groups due to the laser treatment is responsible for the appearance of this additional peak.

In addition, the fitted $Ti_3C_2T_x$ structure has a c parameter of about 20 Å, which corresponds to an interlayer spacing of about 10 Å between the $Ti_3C_2T_x$ nanosheets.[33] X-ray diffraction (XRD) data of our $Ti_3C_2T_x$ films (Figure S1) indicates a larger *c* parameter of ca. 30 Å, corresponding to an interlayer spacing of about 15 Å. However, the interlayer distance in MXenes depends on the nature of the species (solvent molecules and Li salts used during the material synthesis) that are intercalated in between the nanosheets and also on the drying temperature.[35] We argue that the removal of intercalated water in the high vacuum of our experiments (<$10^{-6}$ mbar) and/or the electron beam irradiation most likely account for the measured decrease of interlayer distance.

Together, the SAED and ePDF analysis of the pristine and the laser-processed materials reveal that the medium-range and long-range atomic order of the $Ti_3C_2T_x$ structure is not affected by the laser exposure. The laser processing induces a morphological restructuring of $Ti_3C_2T_x$ nanosheets that diminishes the preferential orientation and results in an SAED pattern that is comparable to the one of the pristine material reordered at −15° tilt angle (compare Figures.3b and 3c). In addition, the laser irradiation alters the short-range order and brings about a new atomic distance at ca. 1.7 Å.

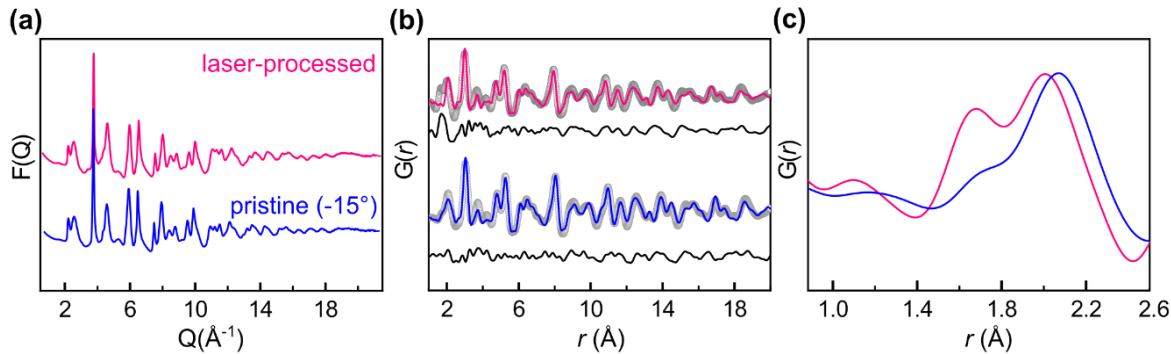

**Figure 4.** ePDF analysis. **(a)** Reduced structure factors, F(Q), of the pristine material, recorded at −15° tilt angle (blue) and the laser-processed material at no tilt (magenta). **(b)** ePDF fits. Light gray circles correspond to the measured data, magenta and blue solid lines are the refined PDFs, and the black solid lines are the residuals. **(c)** Zoom into the low-$r$ ePDF region, showing a characteristic double peak of the Ti-$T_x$ distances of the laser-processed sample (magenta).

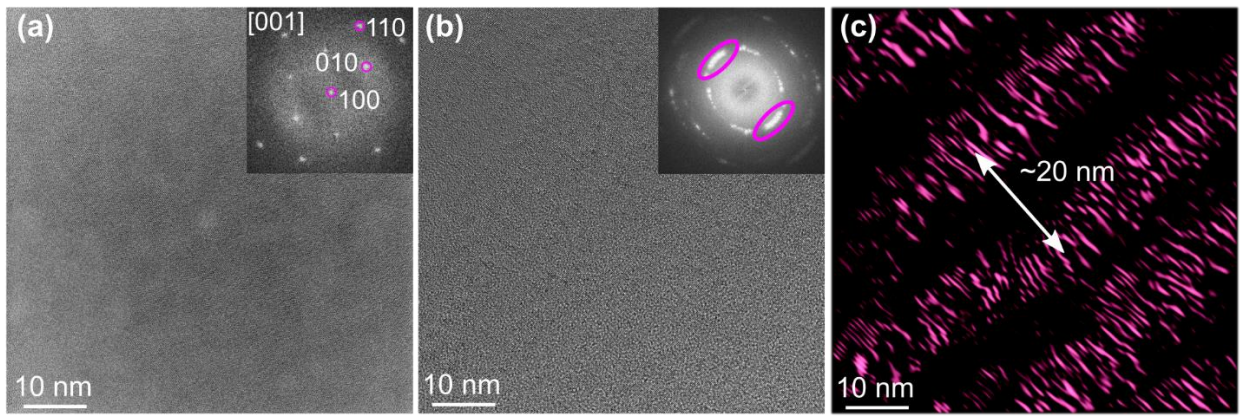

**Figure 5.** Transmission electron microscopy analysis. **(a)** HRTEM micrographs of the pristine material. **(b)** HRTEM micrographs of the laser-processed $Ti_3C_2T_x$ nanosheets. Fourier-transform (FT) patterns are shown as insets. **(c)** Filtered real-space image obtained by selective inverse Fourier transform of the ($10\ell$) arcs marked by magenta ovals in the FT pattern of the HRTEM micrograph in panel **(b)**.

**High-resolution transmission electron microscopy (HRTEM)**

To assess the changes between the pristine material and the laser-processed material at the atomic level, we use high-resolution transmission electron microscopy (HRTEM). Figure 5a shows the HRTEM image and the corresponding Fourier transform (FT) pattern of the pristine material, revealing single-crystalline $Ti_3C_2T_x$ nanosheets that are oriented along the [001] zone axis. The FT pattern of the real-space image (Figure 5b) of the laser-processed $Ti_3C_2T_x$ displays the $(10\ell)$ arcs (Figure 5b, inset). In order to reveal the atomic-scale origin of the diffraction arcs in real space, we calculate the inverse FT image of only the $(10\ell)$ arcs in the computed diffraction pattern. The results are depicted in Figure 5c and highlight only those parts of the lattice that contribute to the creation of the anisotropic diffraction arcs. We see atomic fringes in form of a pattern with a regular periodicity of about 20 Å. This result means that the laser treatment induces undulations (i.e. ripples) of the MXene sheets with a specific periodicity that is much smaller than the laser wavelength (1028 nm). As a result, the originally flat morphology of $Ti_3C_2T_x$ nanosheets on the nanoscale[15] transforms to a regularly rippled hill-and-valley morphology, thereby allowing the $(10\ell)$ reflections to appear in the SAED patterns (Figure 3b) and in the femtosecond data of Figure 1c.

**DISCUSSION**

We first discuss the femtosecond results and afterwards the ripple formation. The observed ultrafast lattice response time of 230 fs is significantly shorter than the rate of several picoseconds that has been suggested before.[36–39] In particular, optical pump-probe investigations of the photo-induced dynamics in $Ti_3C_2T_x$ have previously estimated an electronic response of 100-400 fs and a subsequent lattice heating within only 1-4 ps.[36–39] Our data overturns this picture and directly shows by real-space observation of the atomic dynamics that the lattice becomes hot within merely 230 fs after laser excitation. This value is, to our knowledge, the fastest reported Debye-Waller response of a simple crystal without a phase transition, i.e. excluding the sub-100-fs decay of the Peierls distortion in antimony[40] and the charge-density waves in bismuth[41]. The ultrafast electron-phonon coupling of our MXenes is a signature of an efficient energy and momentum transfer between the electronic band structure and the phonons, similar to the strongly coupled in-plane phonons in graphite[42,43] and the efficient plasmon-phonon coupling in graphene.[44] The ultrafast decay of optical excitations into phonons in $Ti_3C_2T_x$ MXenes might therefore be related to the highly efficient light-to-heat conversion of this material.[17]

The discovered formation of nanoripples in our $Ti_3C_2T_x$ sheets has features and excitation conditions that relate to the laser-induced periodic surface structures (LIPSS) that can be formed on the surface of metals and semiconductors under near-breakdown femtosecond irradiation.[8] For example, sub-wavelength structures on titanium can become as small as 50 nm, as well as aligned with the laser electric field.[45] An elusive feature of this effect is the short wavelength of the fabricated nanostructures that can have dimension much smaller than the optical wavelength. Also in our $Ti_3C_2T_x$ sheets, the periodicity is 50 times smaller than the wavelength. It has been hypothesized that the extremely small modulation period of deep-subwavelength nanoripples in semiconductors and metals is caused by localized surface plasmon resonances under femtosecond nonequilibrium.[46] Our results support this inference. The femtosecond lattice response shows no anisotropy (Figure 1d, inset), suggesting that nanoripple formation is a multi-pulse effect. Yet the final nanoripple orientation is stable against a change of laser polarization, indicating the self-sustaining nature of nanoripples. The threshold character of ripple formation further signifies the role of optical nonlinearities in the process. We suggest that the plasmonic properties of $Ti_3C_2T_x$ [47,48] and its extremely fast lattice reaction time (Figure 1) combine to play an important role in the observed sub-wavelength nanorippling. MXenes are plasmonic materials[47] in the near-infrared region, but behavior of MXenes is different from classical plasmonic metals[48] in that the sheets are isolated and behave like independent 2D structures. The dependence of the laser threshold value on a particular sample spot (±0.5 mJ/cm$^2$) indicates the role of local optical resonances as triggers for the necessary symmetry breaking and self-organization.

The ripples observed in Figure 5c are similar to the topological defects in graphene, where the presence of pentagonal units in hexagonal graphene lattice induces buckling.[49] Based on this analogy and in agreement with the ePDF analysis, where the additional low-$r$ component of the Ti-O/F/C atomic distance has been observed, we conclude that our laser irradiation of an initially flat MXene sheet induces ripples and regular defects that are associated with low-coordinated Ti atoms due to the curvature of the ripples that are formed. There is no laser-induced crystallization or bond-interchange[50,51] nor a phase transition or a chemical reaction, in agreement with theoretical reports that predict that the $Ti_3C_2T_x$ ground state is thermodynamically stable.[52,53] Interestingly, our diffraction data indicates that we produce nano-ripples throughout the entire stack of $Ti_3C_2T_x$ sheets, in contrast to a mere surface-layer modification typically observed in LIPSS. In other words, we modify a bulk stack of nanosheets.

**CONCLUSION**

In conclusion, $Ti_3C_2T_x$ MXene has an ultrafast electron-phonon coupling rate of 230 fs which is by a factor of about ten faster than hitherto supposed. Femtosecond laser treatment of flat $Ti_3C_2T_x$ nanosheets induces a regular hill- and valley morphology in form of nanoripples at a specific and well-defined periodicity that is about a factor of 50 smaller than the laser wavelength. Consequently, $(10\ell)$ Bragg reflections show up in electron diffraction. The orientation of the nanoripples is controlled by the laser polarization and the formation is therefore in part a field effect. A once processed specimen can repeatedly be switched forth and back between a flat and rippled morphology by illumination with laser light. This reversible nanorippling is expected to further enhance the performance of MXenes for various applications.

## METHODS

### Sample Preparation

$Ti_3C_2T_x$ ink (colloidal solution in water, ca. 3 mg ml$^{-1}$) is prepared after chemical exfoliation of the respective MAX precursor, i.e. $Ti_3AlC_2$, in an aqueous mixture of LiF and HCl, following the reported protocol.[35] This ink is drop-casted onto copper TEM grids with 170-μm windows, which forms free-standing thin films after evaporation of the solvent. In few-electron diffraction, the sample must be thin enough to be electron-transparent but thick enough to produce efficient diffraction. To achieve an optimal thickness (as determined experimentally), the concentration of the ink is adjusted using distilled water before drop casting (ca. 1 mg ml$^{-1}$).

### Experimental setup

The ultrafast electron diffraction experiments are carried out on a femtosecond electron beamline (Figure 1a) described in detail previously.[27,28] A laser system with pulse repetition rate of 25 kHz (Pharos, Light Conversion Inc.), center wavelength of 1028 nm and 270-fs pulse duration drives the electron pulse generation, the terahertz (THz) compression, the THz streaking and the sample excitation. A symmetric layout of THz generation for the compression and streaking stages allows to minimize the timing jitter of THz generation. The electron pulses are produced via two-photon photoemission from a back-illuminated gold cathode.[30] A compact electron gun[54] accelerates the electrons to 70 keV, minimizing the influence of vacuum dispersion on the electron pulse duration.[29] We avoid space-charge effects by using less than 8 electrons per pulse.[29] The electron beam is focused with a magnetic lens, and a single-cycle terahertz pulse compresses the electrons down to 75 fs (full width at 1/e$^2$ maximum) on aluminum foil.[27] The electron pulse duration is characterized via THz streaking in a bow-tie resonator[27], providing a temporal resolution of 0.17 pixels/fs (Figure 1b). Streaking broadens the electron beam from 24.8 pixels to 27.4 pixels (full width at 1/e$^2$ maximum), yielding an electron pulse duration estimate of 75 fs full width at 1/e$^2$ maximum or 44 fs full width at half maximum.

The near-infrared (NIR) pump pulses are produced via double-stage second-harmonic non-phase-matched broadening in self-defocusing $\chi^{(2)}$ medium.[55–57] A frequency-resolved optical gating measurement (FROG) yields 55-fs pulses, accounting for the dispersion of the entrance window into the vacuum chamber. Laser fluence and intensity are determined by imaging the actual beam profile in the focus in the vacuum chamber onto a CMOS camera with no assumptions on the beam shape. If a Gaussian fit of the beam profile is applied, it yields a diameter of 208 μm in the focus (full width at $1/e^2$ maximum). The optical beam is incident on the front surface of the MXene-covered copper support grid. The electron beam is focused with a magnetic lens to a size at the sample of 90 µm (full width at $1/e^2$ maximum), measured with a knife-edge scan and the electron camera.

The diffraction pattern is focused and magnified onto the sensor (TemCam F416, TVIPS GmbH) with a second magnetic lens in order to improve the signal-to-noise in the weak diffraction rings while avoiding camera saturation by the zero-order Bragg spot. Astigmatism of the magnetic lens causes a slight distortion of the diffraction rings into ellipses. We calibrate the image rotation and the diffraction angles with a crystalline silicon sample. In the femtosecond pump-probe experiments, fluctuations of pulsed electron current are eliminated by normalizing each evaluated diffraction intensity to the intensity of the direct beam (000 order). The incident average laser power is 7 mW, the pulse energy is 0.28 µJ, the peak intensity is 16.7 GW/cm$^2$ and the fluence is 0.9 mJ/cm$^2$.

The transmission electron microscope is a double Cs-corrected (TEM&STEM) scanning transmission electron microscope (JEM-ARM300F Grand ARM, JEOL Inc.) that was operated at an electron energy of 300 kV. For electron microscopy, samples were prepared by drop deposition of $Ti_3C_2T_x$ colloidal solution onto a copper grid of 200 mesh coated with lacy carbon. All electron diffraction patterns were recorded on a charge-coupled device camera (One View, Gatan Inc.). Azimuthal integration of two-dimensional electron diffraction patterns and their conversion into the reduced pair-distribution function $G(r)$ was done using the freely available eRDFanalyzer software.[58] Structural refinement of the obtained ePDFs is made with the PDFgui software.[59]

A $Pd_{60}Au_{40}$ cross-grating sample serves for calibrating the camera length and instrument resolution parameters $Q_{damp}$ and $Q_{broad}$. The refined values of $Q_{damp}$ (0.044 Å$^{-1}$) and $Q_{broad}$ (0.07 Å$^{-1}$) are fixed during refinement. The camera length is set to reach a reasonable Q value of ca. 22 Å$^{-1}$.

## ASSOCIATED CONTENT

**Supporting Information**. Additional XRD and SAED characterization data, in situ heating and laser switching experiments.

## AUTHOR INFORMATION

**Corresponding Authors**

Mikhail Volkov - University of Konstanz, 78464 Konstanz, Germany; https://orcid.org/0000-0003-1594-8542; Email: mikhail.volkov@uni-konstanz.de

Elena Willinger - ETH Zürich, 8092 Switzerland; https://orcid.org/0000-0001-5117-923X; Email: elenawi@ethz.ch

**Author Contributions**

The manuscript was written through contributions of all authors. All authors have given approval to the final version of the manuscript.

**Funding Sources**

We acknowledge the European Research Council (ERC) under the European Union's Horizon 2020 research and innovation program (grant agreement No. 819573) for partial funding of this work.

**Acknowledgement**

The authors thank F. Krausz for laboratory infrastructure and the Scientific Centre for Optical and Electron Microscopy (ScopeM) of ETH Zürich for providing access to electron microscopes.

Supporting Information for

# Photo-switchable nanoripples in $Ti_3C_2T_x$ MXene

*Mikhail Volkov[1,2,*], Elena Willinger[3,*], Denis A. Kuznetsov[3], Christoph R. Müller[3], Alexey Fedorov[3] and Peter Baum[1,2]*

[1] University of Konstanz, Universitätsstraße 10, 78457 Konstanz, Germany

[2] Ludwig-Maximilians-Universität München, Am Coulombwall 1, 85748 Garching, Germany

[3] Department of Mechanical and Process Engineering, ETH Zürich, Leonhardstrasse 21, 8092 Zürich, Switzerland

**\*Authors to whom correspondence should be addressed:**

mikhail.volkov@uni.kn

elenawi@ethz.ch

## 1. X-ray diffraction characterization (XRD)

The XRD pattern of the initial MXene material $Ti_3C_2T_x$ is shown in Figure S1. The characteristic (002) peak, located at around a diffraction angle $2\vartheta = 6°$, corresponds to an interlayer spacing of 14.8 Å, indicating a unit-cell dimension with parameter $c$ of 30 Å.

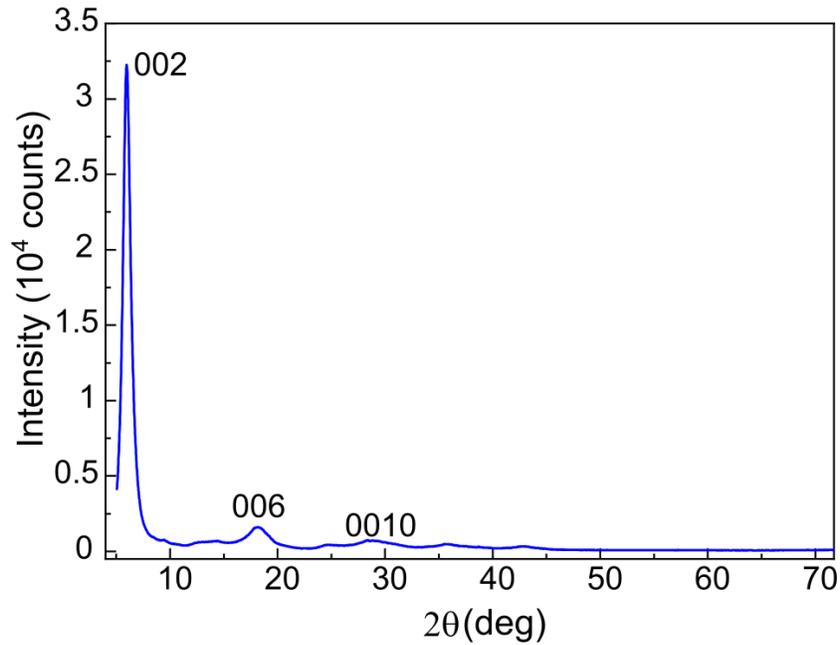

**Figure S1.** X-ray diffraction pattern (XRD) of the pristine MXene material $Ti_3C_2T_x$.

## 2. Selected area electron diffraction (SAED)

Figure S2 shows electron diffraction in form of a tilting series of $Ti_3C_2T_x$. With increasing tilt angle (see the legend to the right), the (10ℓ) peak intensity increases. At the same time, the (110) peak intensity decreases.

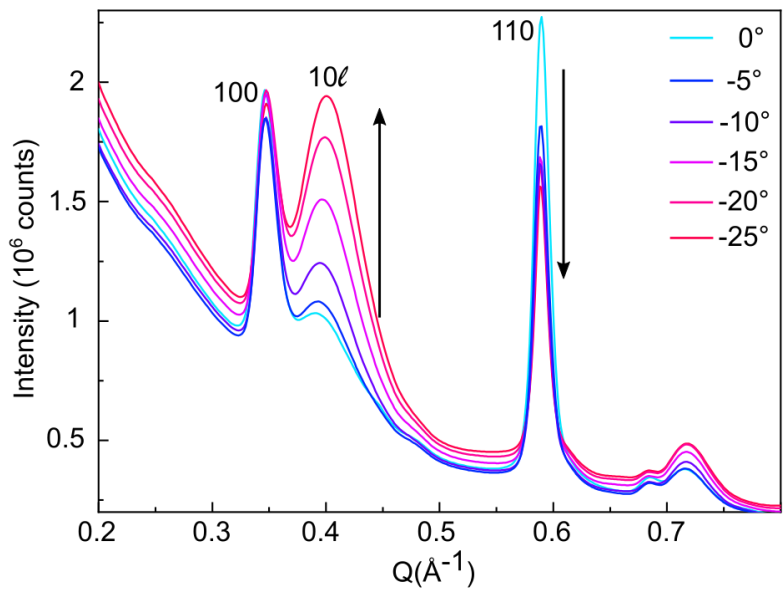

**Figure S2.** Series of SAED patterns of $Ti_3C_2T_x$ as a function of tilt angle (see the inset for color codes and angles used).

## 3. Switching property

$Ti_3C_2T_x$ flakes lose their photo-switching property if exposed to a fluence approaching the breakdown. This regime is distinguished by general loss of electron counts in all diffraction rings (Figure S3a, red-shaded region) above 1.3 mJ/cm². Figure S3a shows intensity in the diffraction rings and arcs after exposing the samples to increasing laser fluence, in a room-temperature sample (laser off). The threshold character of the arc formation is evident (deep blue line).

The opposing dynamics of (110) and (10ℓ) peaks as a reaction to switching the laser on/off is evident from Figure S3b. It shows the differential intensity between laser-hot and room-temperature processed $Ti_3C_2T_x$. While the arcs disappear in the hot phase (blue color) the (110) diffraction ring gains intensity. The slightly elliptical shape of the rings is due to a higher magnetic lens current used in this data set, leading to a more pronounces astigmatism.

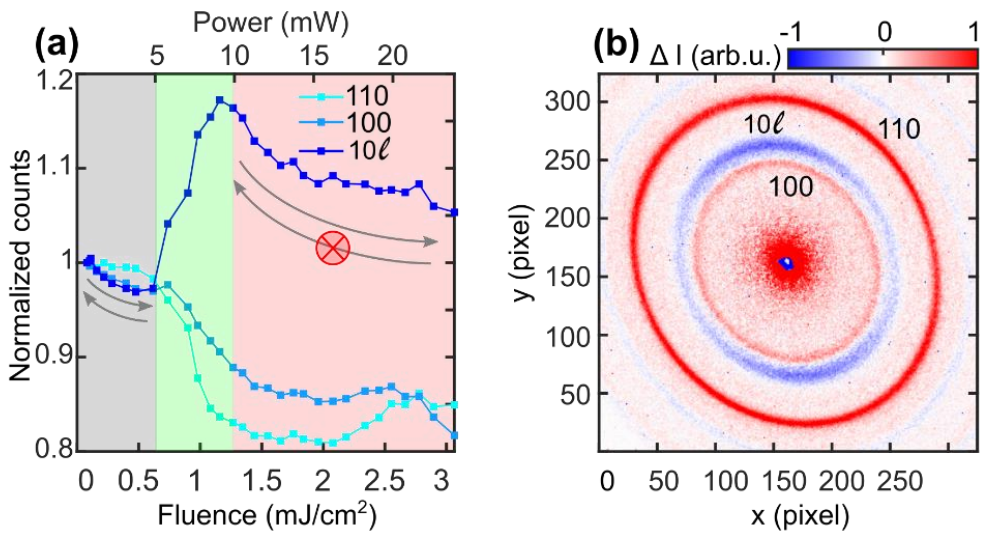

**Figure S3**. **(a)** Electron count in the diffraction rings and arcs of a room-temperature $Ti_3C_2T_x$, exposed to a certain maximum fluence (horizontal axis). Green-shaded area shows the photoswitching range. Red-shaded area indicates the irreversible range with gradual evaporation of the sample. **(b)** Photoswiching dynamics, corresponding to the green area of panel (a). The differential image of the diffraction pattern is "laser on" minus "laser off".

## 4. In-situ heating

Thermodynamic heating, performed in situ within the TEM, does not reproduce the nanorippling effect. This observation is additional evidence that the observed MXene rippling requires high-peak-intensity laser irradiation under involvement of non-equilibrium physics.

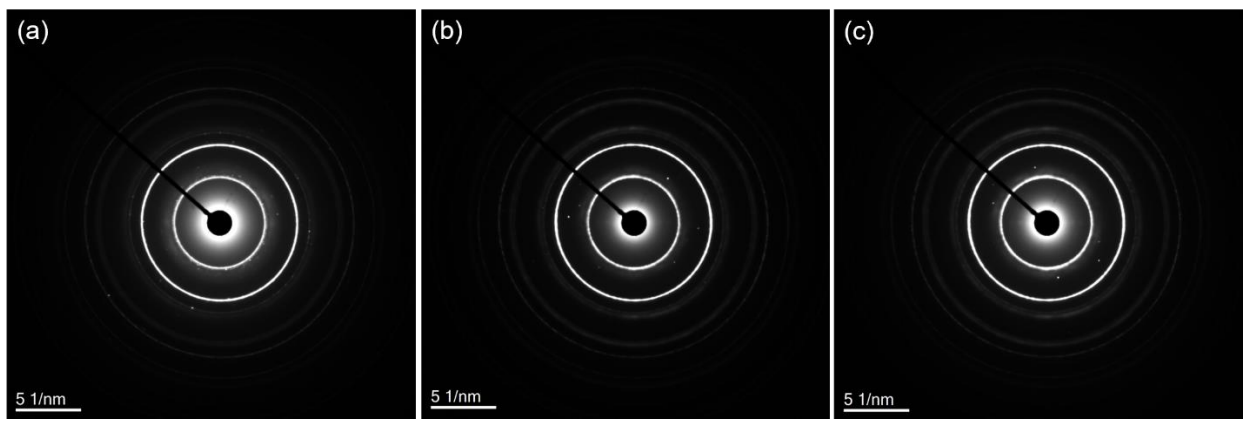

**Figure S4**. SAED patterns during in-situ TEM thermal heating of pristine $Ti_3C_2T_x$. **(a)** Initial diffraction pattern at room temperature. **(b)** Diffraction pattern obtained at room temperature after intermediate thermal heating up to 500°C. **(c)** Diffraction pattern at 500°C. Neither of these patterns shows any indications of arc formation.

25